Ethology of Latent Spaces[1][2]

Philippe Boisnard
Paris 8 — Paragraphes-CITU

***Abstract***

This study questions the presumed neutrality of latent spaces in vision-language models (VLMs) by proposing an ethological approach to their algorithmic behaviors. Contrary to the hypothesis of a homogeneous indeterminacy of these spaces, our analysis shows that each model develops a specific *algorithmic sensitivity*—a differential regime of perceptual salience inherited from its training data, architectural design, and optimization processes.

Through a comparative analysis of three models (OpenAI CLIP, OpenCLIP/LAION, and SigLIP) applied to a corpus of 301 artworks (15th-20th centuries), we demonstrate significant divergences in the attribution of political and cultural categories. Using bipolar semantic axes inspired by vector analogies (Mikolov et al., 2013), we show that SigLIP classifies 59.4% of the artworks as politically engaged, compared to only 4% for OpenCLIP/LAION —African masks receiving the highest scores in SigLIP, while being classified as apolitical by OpenAI CLIP. On the aesthetic/colonial axis, the discrepancy between models reaches 72.6 percentage points.

We introduce three operational concepts: *computational latent politicization*, the process through which political categories emerge without intentional encoding; *emergent bias*, irreducible to statistical or normative biases and diagnosable only through contrastive analysis; and three algorithmic scopic regimes-entropic (LAION), institutional (OpenAl), and semiotic (SigLIP)-characterizing distinct modes of structuring the visible.

Drawing on Foucault (archives), Jameson (*ideologeme*), and Simondon (individuation), we argue that training datasets function as quasi-archives whose discursive formations crystallize within latent space.

This research contributes to a critique of the interpretive conditions of possibility of VLMs when applied to digital art history, and calls for a methodology that integrates the analysis of learning architectures into any delegation of cultural interpretation to algorithmic agents.

***Keywords***

CLIP, latent spaces, emergent bias, latent politicization computational, ethology, digital art history, algorithmic scopic regimes

In 2009, in response to the rapid expansion of digital data production on the internet, Manovich (2009) questioned the possibilities and transformations that might arise from using algorithmic systems to process this growing volume of information:

> "The ubiquity of computers, digital media software, consumer electronics, and networks has led to a massive increase in the number of cultural agents around the world and in the media they create— making it difficult, if not impossible, to achieve a deep understanding of global cultural dynamics using the theoretical tools of the twentieth century. But what happens if we use precisely these developments-computers, software, 'born-digital' cultural content—to track global cultures in new ways?"

---

[1] This article is based on my presentation "Ethology of Latent Spaces" at the Hyperheritage International Symposium (Cultural Heritage in the Age of AI) https://hyperheritage.org/programme-HIS9.pdf. A French version will be published in the symposium proceedings in 2026 by Europa Editions.

[2] All the raw data used for the analysis in this article are available on Zenodo. https://zenodo.org/records/18165882

This question implies a dual shift, affecting both the perception of images and the conceptual frameworks used to understand them, moving from human agency to machine agency. Through this concluding question, Manovich lays the groundwork for a reflection on a fundamental transformation of visual culture, suggesting that perception and interpretation could be delegated to algorithmic agents, rather than remaining exclusively attributed to human consciousness.

Since the connectionist turn marked by the large-scale successes of deep learning (particularly following ImageNet and AlexNet) the systems used for visual analysis and generation have been profoundly reconfigured. They no longer rely solely on explicit descriptors, but on distributed representational spaces learned from large-scale corpora (Krizhevsky et al., 2012; Cardon et al., 2019).

In this context, CLIP-type models (Contrastive Language-Image Pre-training) constitute a methodological rupture (Radford et al., 2021): by jointly learning text and image embeddings, they establish a latent space in which metric similarity becomes a semantic operation —not a term-to-term correspondence, but a network of vectorial compatibilities. This equivalence between metric similarity and semantic similarity constitutes both the founding postulate-and the blind spot—of CLIP models.

It is therefore necessary to distinguish two levels of description of the latent space.

The mathematical latent space is an n-dimensional vector space (typically 512 or 768 dimensions for CLIP) endowed with a metric (generally cosine similarity). The operations performed within it are purely geometrical: projection, distance, interpolation. At this level, nothing is "semantic" — there are only vectors and angles.

The semantic latent space is an interpretation of this mathematical space, through which geometric proximities are read as proximities of meaning. This interpretation is not arbitrary: it is constrained by contrastive learning, which optimizes the alignment between image embeddings and embeddings of texts describing those images. However, it remains an interpretation-and it is precisely here that biases are located.

The transition from mathematical space to semantic space is not a deduction; it is a statistical induction stabilized through training. When we say that two images are "close" in latent space, we are stating a geometrical fact; when we say that they are "similar," we are formulating a semantic hypothesis grounded in the regularity of learned associations. Yet this regularity is not universal—it depends on the training distributions.

Our method exploits precisely this gap: by projecting bipolar semantic axes ("apolitical neutral" → "political engaged") into latent space, we transform a question of meaning into a question of geometry.

The inter-model divergences we observe reveal that the same geometry (projection onto an axis) produces different semantics depending on the model—evidence that the metric/semantic equivalence is locally unstable and corpus-dependent.

Accordingly, what is not specified in a prompt or a query is not "absent": it is precisely what the latent space tends to complete through its learned neighborhoods, producing an algorithmic factuality of images and descriptions-that is, a completion of visual reality based on distributions learned within latent space.

This makes it possible, for example, to grasp the counterfactual outcome produced by Gemini, which generated Nazis characterized as people of color (Huang et al., 2025). As soon as the prompt does not include certain attributes, a non-intentional completion occurs. The representation is completed according to the model's own vectorial equilibria, thereby revealing the implicit hierarchies of its latent space. This dynamic of biased completion was documented as early as 2021 by OpenAI researchers themselves: Agarwal et al. (2021) showed that CLIP encodes stereotypical associations between visual categories and social attributes, and that these biases propagate through usage without explicit intention.

Is there not, then, a blind spot for users of AI models, who believe they control generation without perceiving the internal structuring modalities of the model? The model's latent cultural space—far from being neutral, objective, exhaustive, or indeterminate-rests on semantic polarities and distributions of meaning that are inaccessible to the human agent, due to the functional opacity of the black box and the inaccessibility of the internal hierarchies of latent space.

To further investigate these questions, we employ a computer vision (CV) software tool that makes it possible to understand, in a precise manner, what an AI perceives according to different CLIP models. In doing so, we align ourselves with the perspective of an ethology of algorithmic processes related to latent spaces, opened in 2019 by the Manifesto on Machine Behaviour published in Nature (Rahwan et al.,2019), which itself extends the line of questioning initiated by Herbert Simon (1996): "Natural science is knowledge about natural objects and phenomena. We may ask whether there can also be an 'artificial' science—knowledge about artificial objects and phenomena." This implies no longer generalizing these processes under the generic term "Al," but rather interrogating each model according to its own specificity.

To situate this work within the field of the digital humanities, we pursue the question of a digital art history as proposed by Impett and Offert (2023). Through this inquiry arises the possibility of grasping the issues of perception and interpretation of a visual field by means of AI processes. However, the objective is not merely to apply models to artworks, but to critique the interpretative conditions of possibility inscribed within their representational spaces.

Using a corpus of 301 artworks (15th-20th centuries, including non-exclusively Western productions), we establish a protocol of exploration based on semantic axes (bipolar tensions) and neighborhood visualizations in order to identify inter-model curvatures and asymmetries: (1) variations in fundamental visual properties, (2) variations in salience in object recognition, and (3) variations in the attribution of abstract concepts (including the qualification of an artwork as "political").

In this respect, our investigation aligns with the recent observation that the perceptual alignment of these models-particularly in culturally embedded domains such as art-remains not only insufficiently characterized, but moreover misaligned with our own perception and our own semantic approach. As Andrea Asperti et al. (2025) state: "Yet, despite its ubiquity, the nature and limits of CLIP's perceptual alignment remain underexplored. Although CLIP is optimized to match images with descriptive captions, it is less clear whether this alignment truly reflects human perception-particularly in complex, subjective, or culturally embedded domains such as art."

## *1. Variation of Latent Space Fields*

Understanding latent spaces-that is, n-dimensional spaces composed of vectors-is the fundamental stake of any attempt to grasp what might be conceived as artificial semantic intellection in AI systems. To address this, we employ a software tool that uses the T-SNE reducer (t-distributed Stochastic Neighbor Embedding). While these visualizations do not provide access to the global geometry of latent space, they nonetheless allow for the examination of local neighborhoods and relative tensions between representations (Van der Maaten & Hinton, 2008; Wattenberg et al., 2016). T-SNE produces representations in which data proximity and neighborhood relations are preserved, while the global structure is lost-unlike UMAP-based reductions.

Although these visualizations are not genuine geometric projections, they nonetheless function as heuristic instruments, enabling the identification of recurrent local asymmetries between models. The repetition of these asymmetries across different degrees of semantic axes suggests the existence of distinct perceptual regimes (Wattenberg et al., 2016; Kobak & Berens, 2019). The 2D representations used here thus visualize semantic tensions (vectors) within a two-dimensional reduction of an n-dimensional latent space.

Our method of bipolar semantic axes follows the line of work initiated by Mikolov et al. (2013) on vector analogies. Mikolov demonstrated that embedding spaces encode semantic relations in the form of directions: the vector "king -> queen" is approximately parallel to the vector "man -> woman." Analogously, we construct axes by computing the difference between embeddings of opposing poles ("apolitical neutral" and "political engaged"), and then projecting images onto this direction. The position of an image along the axis measures its relative proximity to the two poles-not a binary classification, but a gradient of salience.

What is of interest to us, however, is how these vectors are mobilized from within a single field of images— here, a corpus of 301 artworks spanning from the 15th to the 20th century, and not exclusively Western.

What we aim to grasp from a decorrelationist perspective is the manner in which the relations established within a model's latent space are the specific result of the model itself. Although this space is indeed related to our own representations-inasmuch as it is structured according to a CLIP process-it nonetheless institutes rules that are heterogeneous with respect to human representation.

In this sense, we adopt a perspective close to that articulated by Jean-Michel Durafour (2018), when he writes: "In correlation, one of the two terms cannot be thought without the other. By contrast, relationism, by affirming that any being exists only insofar as it is in relation (...) also makes it possible to think that there may always exist other relations for a term than the one it has with another term, and therefore that nothing as closed and unidirectional as correlation—particularly the human-object correlation—can impose its laws alone. One can enumerate other relations, between objects, between images, without passing through the constitutive mediation of the human (…..)".

This point is crucial and was already highlighted by Lev Manovich (1998), who articulated the difference between the arrangement of information within a narrative for humans and the ontological nature of the database. Meaning, therefore, is not the result of a pre-conceived and intentionally human architecture, but rather emerges from latent space (Bengio & Vincent, 2013), insofar as latent spaces organize factors of semantic variation whose heuristic rules are drawn from data, architecture, and learning.

The models we examine are:

— OpenAI CLIP Large (ViT-L/14), trained on WIT (WeblmageText), a proprietary dataset of approximately 400 million image-text pairs, curated according to undisclosed criteria (Radford et al., 2021).

— OpenCLIP ViT-L/14, trained on LAION-2B, a subset of the open LAION-5B dataset, which contains 5.85 billion image-text pairs extracted from the web through automated scraping (Schuhmann et al.,2022) Unlike WIT, LAION-5B applies minimal filtering: image-text pairs are retained if their CLIP similarity score exceeds a given threshold, without human curation or semantic verification.

— SigLIP Large, trained on WebLI, a proprietary Google dataset whose exact size and composition have not been disclosed, but which is distinguished by a pairwise learning architecture rather than a contrastive one (Zhai et al., 2023).

This diversity of sources —a curated proprietary dataset (OpenAl), an uncurated open dataset (LAION), and a proprietary dataset coupled with a distinct architectural paradigm (SigLIP)-constitutes precisely the ground of our comparative analysis.

When considering a representation, it is necessary to distinguish three levels of image reality:

1.A first dimension corresponding to the pixel level, for instance chromatic fields, luminance, compositional structures (rule-of-thirds lines, golden ratio lines), or body density.

2. A second level corresponding to the objects contained within an image, such as faces or types of architecture.

3. A third level—the one that concerns us most directly-that of meaning: for example, what does it mean for an image to be politically engaged?

From the very first level of analysis, it is possible to observe that there exists a form of difference in *algorithmic sensitivity* among the three models under consideration. Following Gibson (1979), we hold that perception can be conceived independently of any symbolic or linguistic operation. This hypothesis makes it possible to approach algorithmic perception not as a conscious interpretation, but as a regime of relational salience structured by the properties of the latent space. Only the effort of a non-anthropologizing phenomenological approach can allow us to grasp the characteristics of this *algorithmic sensitivity*. We thus define *algorithmic sensitivity* as the set of differential regimes of perceptual salience through which models implicitly hierarchize visual or semantic properties, according to their training distributions, their architecture, and their optimization processes.

It is nevertheless necessary to clarify our terminological usage. We distinguish three levels of operation:

— Classification refers to the mathematical operation by which a model projects an input (image, text) into a vector space and computes metric proximities. It is a formal operation, entirely describable in terms of functions and parameters.

— Algorithmic perception designates the regime of differential salience by which certain properties of the input become operative in classification while others are ignored or attenuated. In the Gibsonian sense, it corresponds to the extraction of computational affordances-not what the image "means," but what it "affords" within the latent space. This perception is not conscious, but neither is it arbitrary: it is structured by learning.

— Interpretation would designate the attribution of an intentional meaning, the understanding of a signification.

We reject this term for CLIP models: they do not understand, and they do not aim at meaning. However— and this is the core of our analysis—their classifications nonetheless produce interpretative effects for the human agents who use them. The model does not know that a given artwork is "political" or "apolitical," but its classification induces a political reading on the part of the user.

This approach extends Hito Steyerl's (2012) reflection on the status of images in the digital age: if images are no longer made to be seen but to operate, then their analysis can no longer be limited to a hermeneutics of the visible. It must instead become a critique of the invisible operations that structure the perceptual field.

Whereas luminance analysis can be conducted using algorithmic approaches such as CIELAB —which objectively dissociate luminance from chrominance and yield strictly mathematical results-for CLIP models we introduced an objective vector enabling the models, with respect to the 301 images in our dataset, to establish a hierarchy of luminance: "deep shadow underexposed darkness" → "bright overexposed highlight luminosity." This analysis shows that the models are not sensitive to the same aspects of light within an image.

The analysis of the luminance axis reveals that even for a visual property assumed to be objective, the three models produce radically divergent classifications (Figure 1). OpenAI CLIP classifies 97.7 % of the artworks toward the bright pole, compared to 77.2 % for SigLIP-a 20-point gap that attests to distinct algorithmic sensitivities (Figure 2). Even more significant, OpenAI places Joseph Kosuth's conceptual neon works among the most "luminous" images in the corpus (scores ranging from +0.129 to +0.105), despite the fact that these works are photographed against a black background and objectively contain very few bright pixels (Figures 3 and 4). This confusion between the concept of light (the word "neon" in the training captions) and the perception of luminosity exemplifies how latent space encodes semantic associations rather than physical properties.

By contrast, SigLIP correctly identifies the masters of chiaroscuro-Rembrandt, La Tour, Vermeer-as the darkest images, demonstrating a more perceptual sensitivity to pixel-level ratios. LAION CLIP, for its part, overwhelmingly classifies Meidner and his apocalyptic visions toward the dark pole (six works in the top ten), conflating emotional tonality with physical luminosity. These results confirm that *algorithmic sensitivity* is never purely perceptual: it is always already mediated by the discursive formations crystallized in the dataset-the word "neon" activating "light" in OpenAI CLIP, the word "apocalypse" activating "dark" in LAION CLIP-whereas SigLIP, with its pairwise architecture, appears to extract affordances (Figure 5).

This difference arises from the data used to train the model and from the manner in which those data are processed. Training induces a form of relative sensitivity to luminance that stems from what we call a semantic-eidetic culture. By model culture, we do not designate an intentional symbolic system, but rather the aggregated effect of statistical regularities. This, in turn, induces a form of computational culture defining a non-conscious algorithmic intentionality, that is, a non-reflexive functional orientation of the algorithm. Each model thus comes to perceive the light of an image differently, giving rise to a specific form of behavior for each of them, in the sense that "the behavior of a system is implicitly induced rather than explicitly constructed" (Bommasani et al., 2021).

Our use of the concept of emergence requires a distinction between weak (epistemic) emergence and strong (ontological emergence. Weak emergence designates properties that are not explicitly programmed but are, in principle, deducible from the configuration of the system's components-their emergent character lies in our practical inability to compute them, not in an irreducibility in principle.

Strong emergence, by contrast, designates constitutively irreducible properties that could not be predicted even with exhaustive knowledge of the system (Bedau, 1997; Chalmers, 2006). The emergent behaviors of foundation models, in our view, fall under weak emergence that is nevertheless practically irreducible: the biases we identify are, in principle, contained within the network weights and the training distributions, but their combinatorial complexity renders them inaccessible to direct analysis. This is precisely why an ethological approach—the observation of behaviors rather than the deduction of mechanisms—imposes itself as a method. We do not claim that these biases are ontologically unpredictable; rather, we observe that they are epistemically opaque, and that only contrastive experimentation allows them to be revealed.

This position has an important methodological consequence: it implies that *emergent biases* could, in principle, be anticipated or corrected —but at a computational and epistemic cost that exceeds current capacities. Opacity is not an ontological fatality; it is a technical condition that calls for appropriate investigative instruments.

With regard to the second level, concerning the distinction of contents, it should likewise be noted that it would be possible to employ an approach based on YOLO, for example, which-thanks to its regression-based learning and its label-driven image categorization schemes—produces highly precise and determinate results.

The vector we use here is building-landscape → human body-face. The analysis of the object axis ("building-landscape" -> "human body-face") highlights a phenomenon of very strong divergence:

SigLIP recognizes only 6.9% of the images as representing a human body, compared to 59.7% for OpenAI CLIP and 52.1% for LAION — a ninefold gap that reveals radically divergent categorical definitions (illustration 6).

An examination of the extremes shows that SigLIP classifies as "body" only academic or naturalistic nudes — Degas, Otto Müller, Fidus — characterized by anatomical proportions conforming to classical canons (maximum scores of +0.032). By contrast, bodies in modern art — Schiele, Bacon, Bellmer, Kirchner — which are deformed, fragmented, or stylized, are invisibilized by SigLIP, as are African masks, which do not appear in its top 10. OpenAI CLIP and LAION, by contrast, integrate these transgressive representations: Bacon occupies four positions in LAION's top 10, and African masks appear in second position (+0.176).

Even more revealingly, SigLIP classifies Vermeer (The Wine Glass) and Hopper (New York Movie) toward the "landscape" pole, even though these works contain human figures - architectural space taking precedence over the bodies it contains. This bias toward corporeal realism most likely derives from the labeling practices of the WeblI dataset, where modern paintings are captioned as ,"expressionism","modern art," or "painting" rather than "human body." The body of modern art is thus named within the discursive formations of the dataset, producing a systematic invisibilization, which confirms that object recognition is never a neutral operation but always-already a culturally conditioned interpretation, shaped by learned categories.

This is further confirmed when one observes the most sharply contrasted gaps between SigLIP and OpenAI CLIP (illustration 7). SigLIP encodes a corporeal realism bias that systematically invisibilizes non-realist representations of the human body and face.

From this second analytical angle, we can identify the following points:

1. The existence of an *algorithmic sensitivity* specific to each model, a sensitivity that is directly linked to the training data used to construct it and to the model's specific encoding processes. Despite a shared design framework, the different foundation models - here OpenAI CLIP, OpenCLIP, and SigLIP - can manifest divergent perceptual regimes, precisely because their emergent properties are structured differently (Bommasani et al., 2021).

2. It is apparent that SigLIP exhibits a more general, more "public-facing" mode of perception than OpenCLIP or OpenAI CLIP. The significant gap between SigLIP and the OpenAI CLIP and OpenCLIP models stems from a different mode of operation and, consequently, from a specific form of emergence. Whereas CLIP relies on global relations among all pairs within a batch - implying that the geometry of the latent space is governed by the overall dynamics of similarities — SigLIP constructs local, symmetric, pairwise relations. In this sense, CLIP perceives data in a more complex manner, inducing attractors grounded in the global context, whereas SigLIP establishes more localized and stable connections (Zhai et al., 2023).

3. None of the models shares the same sensitivity; from the results, it is possible to infer the foundations of their respective training regimes and, by extension, their cultures. *Algorithmic sensitivity* does not correspond to an explicit normative bias, but rather to a differential regime of perceptual salience, arising from training distributions and optimization functions (Geirhos et al.,2018). Algorithmic behaviors, therefore, are not intentions - neither for human consciousness nor for the models themselves - but emergent properties linked to architecture, data, and optimization objectives (Goyal & Bengio, 2022).

## 2. *From a Political Perception of CLIP Models*

Our third level of analysis, developed in this second part, focuses on more abstract vectors - on ideas.

What we seek to examine is the possibility of a political understanding of their constitution. Where Chatonsky (2022) conceives the latent space as a common that remains politically indeterminate - "The latent space would be a common without a constituted politics, but with a politics forever deferred, remaining possible as if suspended within itself, without before or beyond" — our analysis shows that implicit political regimes can emerge locally, in a differential manner depending on architectures and training distributions.

The problem, therefore, is not so much the absence of politics as the hypothesis of a homogeneous indeterminacy of the latent space as if it were ontologically neutral and non-differentiated across models), a hypothesis that tends to neutralize the local differentiations produced by architectures, datasets, and discursive regimes. The political substrate of the latent space is structured both by the choice of datasets, by the image-text pairs that are constructed, and by the algorithmic encoding processes and optimization functions of the model. As such, from their very inception, these models envelop political potentials that will induce biases.

Human perception, when confronted with the 301 works of art that constitute the dataset, will itself establish a form of political classification, depending on individual differences. It is evident, for example, that a consciousness lacking familiarity with the political and aesthetic stakes of the early twentieth century will not perceive in what sense the works of Russolo carry a political charge. Such a viewer will tend to privilege the explicit referent of representation - for instance, in Otto Dix, the visible presence of war — whereas Russolo will be understood primarily through aesthetic categories (modern art). Thus, without entering into a detailed analysis of human perception, it is clear that the political approach to a work of art by human consciousness relies on a cultural environment, historical and political knowledge, and a symbolic apprehension of representations. As Bourdieu (1979) demonstrated, aesthetic judgment - and by extension political judgment - rests on internalized cultural dispositions that orient perception without being consciously mobilized. There is a passive process of structuring judgment and its enunciation.

Accordingly, the question is not whether an agent - human or artificial - intends to produce a political interpretation, but rather how a given system of representation makes certain classifications possible, probable, or salient, independently of any intention. It is precisely on this terrain that the analysis of AI models we have undertaken is situated: not as political subjects, but as devices for structuring the visible and the sayable, whose interpretative effects emerge from their architectures and their data (Rahwan et al., 2019; Bommasani et al., 2021).

We propose the term *computational latent politicization* to designate the process by which political categories or qualifications emerge within the latent space of an AI model, not as the result of an intention or intentional encoding, but as a structural effect of uneven discursive distributions. Although a number of algorithmic biases and unintentional political effects have already been identified (Noble, 2018; Benjamin, 2019; Berg et al., 2022), no formal term has yet been proposed to systematically analyze the emergence of categories within vector representations. The notion of *computational latent politicization* seeks to fill this gap.

This mechanism extends, within the visual domain, the findings of Caliskan et al. (2017) on textual embeddings: just as implicit associations between concepts ("woman" and "family," "man" and "career") emerge from statistical co-occurrences in corpora, associations between images and political categories emerge from the image-text pairs of CLIP datasets. The crucial difference is that, in the case of CLIP, the bias crosses the boundary between language and vision: an image can be "politicized" not by its intrinsic visual content, but by the captions with which it was associated during training.

It manifests, as we will show, through asymmetries of interpretative salience depending on the cultural data or visual forms represented. Thus, by political, we do not mean an explicit ideology, but rather a regime of differentiation of the visible, through which objects acquire a certain definition and occupy particular positions (relations of conflict or proximity) with respect to other objects (Rancière, 2000). AI models therefore do not learn ideologies, since they do not engage in symbolic reflection; rather, they generate *ideologemes*, that is, structures of meaning that condition any possible ideological articulation.

Data — and discursive data in particular — constitute the raw material of *latent politicization*, in a Foucauldian sense (Foucault, 1969). Our hypothesis is that training datasets function as *quasi-archives*: they do not determine what can be said (the model can generate statements absent from the dataset), but they structure what can become salient, what will be probable, and what will be proximate within the latent space. *Latent politicization* is therefore not the effect of an explicit content of the dataset (which would amount to a statistical bias), but rather the effect of the discursive formation instantiated by the dataset — the regularities, co-occurrences, and systematic exclusions that, once aggregated in the latent space, produce regimes of visibility. In this Foucauldian sense, CLIP models do not learn contents; they internalize rules of formation.

The *ideologeme*, then, is not a constituted ideology, but, in Jameson's sense (1981), a minimal, pre-reflexive unit of meaning, capable of entering into subsequent ideological configurations. Models do not produce ideology; they produce structures of differentiation that condition the very possibility of its emergence.

This characterization calls for a terminological clarification. The literature on algorithmic bias generally distinguishes between two meanings: statistical bias, which refers to a measurable deviation from a reference distribution (for example, the underrepresentation of a group in a dataset), and normative bias, which qualifies a deviation from an ethical or political norm (for example, discrimination within an institutional context). Our concept of *computational latent politicization* falls under neither category. It designates a third type: *emergent bias* — a structural effect that is neither intentionally encoded nor directly traceable to statistical distributions, but which results from the vectorial entanglements produced during the formation of the latent space.

*Emergent bias* cannot be measured by comparison to a "fair" distribution (there is no ground truth for what the political classification of a Vermeer should be), nor can it be evaluated against a pre-established norm (no consensus exists regarding what would constitute a true or objective reading of political engagement in art). It is instead diagnosed through contrastive analysis of inter-model divergences: it is precisely because SigLIP, OpenAI CLIP, and OpenCLIP/LAION produce radically different classifications on the same works that we can infer the existence of distinct regimes of salience, irreducible both to the intentions of the designers and to the raw statistical properties of the datasets. In this sense, the *ideologeme* — a minimal structure of differentiation - constitutes the analytical unit of *emergent bias*.

This distinction extends the call by Blodgett et al. (2020) to clarify the multiple uses of the term bias in AI studies. Our specific contribution is to identify a third type - *emergent bias* - which is particularly salient in multimodal models, where text-image entanglements produce semantic effects that are not reducible to input distributions.

For this part of the analysis, we adopt an approach that is both illustrative and quantitative, enabled by our software. First, in an initial test, we compare the three models along a semantic tension: (apolitical neutral -> political engaged). Our test does not measure a political understanding of the artworks by the models, but rather the way in which discursive categories associated with politics locally structure the latent space.

With the OpenAI CLIP Large model (illustration 8), we observe that the "politically engaged" label is relatively moderate (33.3%). It applies, for instance, to photographs by Larry Clark (score: 0.024), works by Kosuth (+0.027), paintings by Kirchner (0.023), and Otto Dix (+0.027) (illustration 9). Somewhat unexpectedly, paintings by Gauguin and by Picasso from the Cubist period influenced by African masks also appear. Notably, African masks are classified as non-political (-0.028) (illustration 10), as are paintings from Picasso's Rose Period and works by Hopper. What emerges is that OpenAI associates the notion of politics primarily with avant-garde movements, that is, with an institutional, museological, and critical discourse. Moreover, when examining its variance in inter-model correlations, we observe that it remains balanced around zero .

With the OpenCLIP/LAION model, very few artworks are classified as politically engaged (4%). Certain specificities already observed in OpenAI CLIP reappear here (illustration 11). Photographs by Larry Clark, as well as the nudes of Schiele and Kirchner, are among the most significant works in its analysis. What is detected is the transgressive body, nudity, and marginality. Even in these cases, however, the scores remain quite low (Schiele's Self-Portrait in Red Shirt, which receives the highest score, reaches only 0.019)

As we will see with the SigLIP model, an entirely different partition emerges.

A large number of images are identified as political (59.4%) (illustration 8). African masks are strongly associated with political engagement (the two highest scores correspond to African masks: +0.052 and +0.050). They are followed by Meidner and his Apocalypse, Russolo, and Picasso.

SigLIP reveals an asymmetry of salience between "African art" and "political engagement", suggesting that its training distributions favor stereotypes according to which non-Western art is intrinsically "political", whereas Western art is perceived as "neutral" by default (illustration 12). This asymmetry falls under what Mohamed et al. (2020) describe as algorithmic coloniality: the reproduction, within AI systems, of epistemic schemas inherited from colonialism-most notably the systematic reduction of non-Western cultural productions to ethnic or political intentions, while Western productions are essentialized as universal or purely aesthetic.

Several hypotheses can be formulated:

1. Images of African art in the training dataset were likely accompanied by contextual descriptions referencing cultural, historical, or "political" aspects. Such works are frequently framed in Western media and texts through political lenses (colonialism, resistance, cultural identity).

2. Western art, by contrast, is more often described in purely aesthetic or technical terms.

This hypothesis could be further substantiated by examining the curvatures of the latent space, and thus the patterns of proximity and attraction it produces.

We observe that the curvature from Picasso toward African masks is clearly attested at the level of vectorial relations. Very strikingly, however, this curvature is not reciprocal. In this sense, we can clearly identify a form of iconological genealogy that privileges Picasso's works (illustration 13).

Moreover, SigLIP's integration of Expressionist works among the highest scores indicates that it has associated discourses of aesthetic rupture with political rupture in the avant-gardes.

Our second test examines five axes simultaneously in order to confront and consolidate our initial results: (1.Conflict axis: explicit violence (war, weapons, conflicts, protest) / implicit power, social tension, alienation, oppression, inequality; 2.Institution / subversion axis: museum, canon, heritage, masterpiece, tradition / subversion, protest, transgression, critique, resistance; 3.Political aesthetics axis: aesthetic, decorative, formal, timeless, universal / colonialism, domination, exoticism, exploitation, otherness; 4. Body / norm axis: ideal body, beauty, proportion, harmony, anatomy / violence, deformation, suffering, sexuality, trauma; 5. Power axis: order, stability, harmony, balance, tradition, continuity / power, domination, oppression, control, authority, hierarchy.)

As we will see, the synthetic analysis of these five axes confirms our initial, broader and more generic analysis. The most pronounced inter-model divergence appears on the "aesthetic/formal" <—> "colonial/exoticism" axis (illustration 14), where SigLIP classifies only 9.2% of the images toward the "colonial" pole, compared to 69.6% for OpenAI CLIP and 81.8% for OpenCLIP/LAION — an extreme gap of 72.6 percentage points between the two most distant models.

This suggests that SigLIP's training corpus (WebLI) underrepresents postcolonial critical discourse, or that it encodes a distinct semantic cartography of "exoticism" that resists the politicized readings of non-Western artistic production. Conversely, the systematic over-attribution of OpenCLIP/LAION to the colonial pole ($\sigma = 0.047$, the highest variance on this axis) indicates stochastic associations inherited from web-scraped image-text pairs without prior curation.

On the institution/canon <—> subversion/critique axis, SigLIP exhibits the strongest bias toward subversive readings (85.8%), combined with the lowest variance ($\sigma = 0.013$), suggesting a stable integration of the contemporary critical vocabulary of art history. OpenCLIP/LAION, despite a comparable directional bias (72.3%), displays a variance three times higher ($\sigma = 0.070$), indicative of stochastic rather than semantically coherent associations. OpenAI CLIP occupies an intermediate position (74.9%, $\sigma = 0.030$), consistent with training on curated institutional and alternative texts that balance canonical and critical frameworks.

The "ideal body" <—> "violence/trauma" axis reveals that OpenAI CLIP is particularly sensitive to corporeal transgression (87.8% toward the trauma pole), aligning with feminist and psychoanalytic readings institutionalized within museological discourse. SigLIP demonstrates a distinct sensitivity to fetishistic and fragmented bodily representations (Bellmer, Oda Jaune), while the intermediate position of OpenCLIP/LAION (59.7%), coupled with a high variance ($\sigma = 0.046$), suggests an incoherent encoding of somatic semantics.

It is worth noting that the "order/harmony" <—> domination/power axis produces a rare convergence: all three models classify approximately 70-77% of artworks toward the "order" pole. However, a closer analysis reveals a significant divergence with respect to canonical works: The Music Lesson by Vermeer receives opposing classifications from SigLIP (+0.026, toward domination) compared to OpenAI CLIP (-0.066) and OpenCLIP/LAION (-0.083), the latter two privileging harmony. These results are consistent with the hypothesis that SigLIP has internalized critical readings of bourgeois domestic space as sites of gendered power relations, a discourse largely absent from the training distributions of the other models.

Taken together, these results reinforce the hypothesis that biases in vision-language models are not artifacts or errors, but rather discursive formations inherited from training corpora, amenable to an archaeological analysis through contrastive semantic probing, and shaped by both architectural choices and optimization strategies.

### *3. Implication*

The latent spaces of CLIP models are neither neutral nor purely objective. Our analysis shows that, depending on the model, the interpretation of semantic tensions is governed by biases in labeling and data classification within the training models. While latent spaces may at first appear to be defined by objective mathematical tools of data compression and representation-seemingly free of ideological commitments—this neutrality is, in fact, only apparent.

Consequently, the question is not whether an agent-human or artificial-intends to produce a political interpretation, but rather how a system of representation renders certain differentiations possible. This approach resonates with the philosophy of Gilbert Simondon, for whom meaning is not the product of intention but the effect of a process of individuation within a given metastable milieu (Simondon, 1958).

AI models can thus be understood not as political subjects, but as devices in the process of individuation, whose latent space constitutes a preindividual field of semantic potentialities, structured by informational tensions arising from data and architectural constraints.

Variance analysis across all axes consistently positions OpenCLIP/LAION as the most unstable model (mean $\sigma = 0.044$), OpenAI CLIP as intermediate (mean $\sigma = 0.024$), and SigLIP as the most coherent (mean $\sigma = 0.014$). This tripartite distinction can be further illuminated through the concept of scopic regimes proposed by Martin Jay (1988) to characterize historical modalities of vision in modernity. Jay notably distinguishes Cartesian perspectivism (a centered, controlled, objectifying gaze), the Dutch art of describing (a gaze attentive to surfaces, textures, and details without hierarchy), and the Baroque (an excessive, proliferating, decentered gaze).

By analogy, we propose to characterize the three models as instantiating three algorithmic scopic regimes :

— OpenCLIP/LAION exhibits a form of latent entropic politicization, characterized by an accidental politicization inherited from the web. This "noise" is not random: as demonstrated by Birhane et al. (2021), the LAION dataset contains systematically biased distributions that propagate into the latent space in the form of stochastic but non-neutral associations;

— OpenAI CLIP manifests an institutional politicization, a controlled and referential regime that reproduces Western museographic hierarchies;

— SigLIP demonstrates a latent semiotic politicization, projecting a contemporary critical and theoretical vocabulary onto visual forms that exhibit strong internal coherence, albeit with a potentially reductive systematicity.

It is necessary to distinguish several levels in order to understand the political implications of a latent space. The level of the explicit and intentional constitution of semantic biases, while deserving documentation and study, is not the most relevant for our investigation. Indeed, it has been noted for more than a decade that the choice of dataset data and their statistical distributions introduce biases in generative AI systems, whether language-based or image-based. Sambasivan et al. (2021) have shown that such data biases are not isolated accidents but "cascades": technical choices (annotation formats, filtering criteria, scraping sources) propagate throughout the pipeline and produce cumulative effects.

However, there exists a deeper level of bias production-what we have termed *emergent bias*. Unlike statistical biases (traceable within datasets) and normative biases (evaluated against external criteria), *emergent bias* forms within the a priori non-measurable entanglement of vectorial relations that occur during the structuring of the latent space. It can only be detected a posteriori, through contrastive analysis, and cannot be corrected without transforming the model's architecture itself.

For a digital art history to be meaningfully constituted, it is therefore necessary first to interrogate the models we employ. These biases of meaning-political, ethnic, historical-are more difficult to discern than group or gender representation biases that depend directly on datasets, because they arise from semantic drifts internal to the structuring of a model's latent space, and may emerge without any form of intentionality. This stands in contrast to group-representation biases, which are more directly traceable, as they are linked to the statistical distributions of training data.

From our analysis, we perceive not only that there exists a machine perception, but also that this machine perception-particularly at the level of the models that underpin it-must be understood, model by model, as a specific logic of approaching the objects that are seen.

As stated by Ghosal et al. (2023): "Visual understanding is one of the most complex tasks in artificial intelligence. Among the many challenges associated with it, image question answering has been formulated as a task that tests the ability of a system to understand the elements of an image in a way similar to how humans interact with images." Consequently, a key challenge lies in the design of models that enable a form of visual understanding of the proposed content-not only with respect to the nature of pixels, nor solely to the objects that are described, but more fundamentally to the semantic networks of meaning that are woven into the representation.

There can be no digital art history without an in-depth investigation into the different models and the ways in which they interact with an image at the level of interpretation. Whenever biases are present, they must be made explicit, as they introduce cultural representational distortions depending on the context.

As Impett and Offert (2023) state: "There can be no visual analysis of images using multimodal models that is not also, at the same time, a critique of the conceptual space inherent to the model. Tool and data, in the age of multimodal models, exist in a reciprocal relationship: looking at data with multimodal models means looking at multimodal models with data." However, this critique of the conceptual space does not concern data alone, but also the crossings and entanglements of data involved in the construction of the latent space-relations that are more or less expected, and largely unconscious for the human agent, in contrast to what is enacted through the automated processes of reduction that construct the virtuality of the latent space.

## *4. Limitations*

This study presents several methodological limitations that must be explicitly acknowledged.

— Corpus. Our analysis is based on a corpus of 301 artworks, which is sufficient to identify inter-model trends but insufficient for robust statistical generalization. Although the selection spans five centuries and includes non-Western productions, it remains dependent on our own criteria of constitution, potentially introducing a confirmation bias. Expanding the corpus-particularly toward contemporary art, digital art, and extra-European productions-would allow for testing the stability of our observations.

— Visualization. The t-SNE reductions used for the illustrations preserve local neighborhoods but distort the global geometry of the latent space (Van der Maaten & Hinton, 2008). The "curvatures" we identify should therefore be understood as heuristic indicators, not as exact measurements of vectorial structure.

Complementary analyses using UMAP or direct inspection of cosine distances would strengthen the validity of these observations.

— Semantic axes. The five bipolar axes we defined (conflict, institution/subversion, aesthetic/colonial, body/norm, power) are not orthogonal; their mutual correlations were not systematically measured.

Moreover, the choice of terms constituting each pole (e.g., "colonialism," "domination," "exoticism") encodes our own critical presuppositions. Alternative formulations would likely yield different results— thereby further confirming our thesis regarding the instability of the metric/semantic equivalence.

— Reproducibility. CLIP and SigLIP models are deterministic with fixed architectures; however, the hyperparameters of our visualizations (t-SNE perplexity, number of iterations) were not systematically varied. A sensitivity analysis with respect to these parameters would improve the robustness of the conclusions.

— Causality. Our contrastive method enables the identification of inter-model divergences, but not the precise establishment of their causes. While we infer that the observed differences stem from datasets and architectures, we cannot isolate the respective contribution of each factor without access to proprietary training data (WIT, WebLI) or without controlled ablation experiments.

Nevertheless, these limitations do not invalidate our conclusions but rather circumscribe their scope. We propose an investigative methodology and a set of operative concepts (*latent politicization*, emergent bias, *algorithmic sensitivity*) whose generalizability it be tested across other corpora, other models, and domains beyond the history of art.

## *5. Conclusion*

As Trevor Paglen writes (2016): "This invisible visual culture isn't just confined to industrial operations, law enforcement, and 'smart' cities, but extends far into what we'd otherwise—and somewhat naively— think of as human-to-human visual culture." Our analysis of CLIP latent spaces confirms this intuition: the visual culture of art is now mediated by algorithmic regimes of visibility, whose categories not only escape the consciousness of the human agents who use them, but are also not reflexively apprehensible by the systems themselves.

As human agents increasingly rely on and delegate their processes of reasoning and meaning-making to AI agents, it becomes urgent to critically examine the learning architectures and mechanisms of world-reading embedded in models trained under distinct learning paradigms. The risk lies in invisible biases that may be constitutive of the models themselves, or generated through dimensional entanglements, and that remain imperceptible to users.

Moreover, if we seek to establish the possibility of a digital art history, it must be acknowledged that while human-centered art history has long been haunted by ethnocentric, cultural, historical, and political biases, a digital perspective on art history must critically interrogate the foundational processes of model production on which it relies. Such a critique must address both the image-text datasets and the dimensionality-reduction algorithms used to structure the vector spaces mobilized in analysis.

(Agarwal *et al.* 2021) Agarwal, S., Krueger, G., Clark, J., Radford, A., Kim, J. W., & Brundage, M. (2021). Evaluating CLIP: Towards Characterization of Broader Capabilities and Downstream Implications. arXiv:2108.02818

(Asperti *et al.*, 2025) Asperti A., Dessi L., Wu N., Toneti M.C., Does CLIP perceive art the same way we do? , arXiv:2505.05229v1 [cs.CV]

(Bedau, 1997) Bebau M.A, Weak Emergence. Philosophical Perspectives, 11, 375-399.

(Bengio et al., 2013) Bengio Y., Courville A., Vincent P., Representation Learning: A Review and New Perspectives (2013) arXiv:1206.5538v3 [cs.LG]

(Birhane et al. 2021) Birhane, A., Prabhu, V. U., & Kahembwe, E. (2021). Multimodal datasets: misogyny, pornography, and malignant stereotypes. arXiv:2110.01963

(Blodgett *et al.* 2020) Blodgett S. L., Barocas, S., Daumé III, H., & Wallach, H. (2020). *Language (Technology) is Power: A Critical Survey of "Bias" in NLP*. ACL 2020.

(Bommasani *et al.* 2021) Bommasani R., A. Hudson D., Adeli E., Altman R., Arora S., von Arx S., S. Bernstein M., Bohg J., Bosselut A., Brunskill E., Brynjolfsson E., Buch S., Card D., Chen A., Creel K., Quincy J., Moussa D., Esin D., Stefano D., Castellon R., Chatterji N., Demszky D., Etchemendy J., Fei-Fei L., Finn C., Gale T., Gillespie L., Goel K., Donahue C., Ethayarajh K., Goodman ., Grossman S., Guha N., E. Ho D., Hong J., Jurafsky D., Kalluri P., Khattab O., Kumar A., Hashimoto T., Hsu K., Huang J., Karamcheti S., Henderson P., Icard T., Keeling G., Hewitt J., Jain S., Khani F., Lisa X., Suvir L., Tengyu M., Ali M., Eric M., Zanele M., Pang M., Koh W., Ladhak F., Li X., Krass M., Krishna R., Kuditipudi R., Lee M., Lee T., Leskovec J., Levent I., D. Manning C., Nair S., Narayan A., Newman B., Nie A., Niebles J.C., Nilforoshan H., Narayanan D., Nyarko J., Ogut G., Orr L., Papadimitriou I., Park J.S., Piech C., Portelance E., Potts C., Raghunathan A., Reich R., Rong F., Roohani Y., Ruiz C., Ryan J., Ré C., Sagawa S., Santhanam K., Shih A., Srinivasan K., Ren H., Sadigh D., Tamkin A., Taori R., W. Thomas A., Tramèr F., E. Wang R., Wang W., Wu B., Wu J., Wu Y., Michael S., Michihiro X., Jiaxuan Y., Matei Y., Michael Z.,

Tianyi Z., Xikun Z., Yuhui Z., Zheng L., Zhou K., Liang P., On the Opportunities and Risks of Foundation Models, arXiv:2108.07258v3 [cs.LG]

(Bourdieu, 1979) Bourdieu P., La distinction. Critique sociale du jugement. Minuit.

(Caliskan et al. 2017) Caliskan, A., Bryson, J. J., & Narayanan, A. (2017). Semantics derived automatically from language corpora contain human-like biases. *Science*, 356(6334), 183-186.

(Cardon et Al., 2019) Cardon D., Cointet J.P., Mazières A., La revanche des neurones , L'invention des machines inductives et la controverse de l'intelligence artificielle. https://shs.cairn.info/revue-reseaux-2018-5-page-173?lang=fr

(Chalmers, 2006) Chalmers D.J. Strong and Weak Emergence. In P. Clayton & P. Davies (Eds), The Re-Emergence of Emergence (pp.244-254). Oxford University Press.

(Chatonsky, 2022) Chatonsky, G., L'espace latent comme commun sans communauté, https://chatonsky.net/commun-sans-communaute

(Durafour 2018) Durafour J.M., Cinéma et cristaux : traité d'éconologie, Mimésis, 2018, p.19

(Geirhos *et al.* 2018) Geirhos R., Rubisch P., Michaelis C., Bethge M., A. Wichmann F., Brendel W., ImageNet-trained CNNs are biased towards texture; increasing shape bias improves accuracy and robustness (2018), arXiv:1811.12231v3 [cs.CV]

(Gibson, 1979) Gibson, The ecological Approach to visual perception. Psychology Press Classic Editions https://library.uniq.edu.iq/storage/books/file/The Ecological Approach to Visual Perception Approach/1667383098The Ecological Approach to Visual Perception Classic Edition (James J. Gibson) (z-lib.org) (1).pdf

(Ghosal et al. 2023) Ghosal D, Majumder G., Ka-Wei Lee R., Mihalcea R., Poria S., Language Guided Visual Question Answering: Elevate Your Multimodal Language Model Using Knowledge-Enriched Prompts , arXiv:2310.20159v1 [cs.CV]

(Goyal & Bengio., 2022) Goyal A., Bengio Y., Inductive Biases for Deep Learning of Higher-Level Cognition, arXiv:2011.15091 [cs.LG]

(Huang et Al., 2025 ) Huang J.T, Yan Y., Liu L., Wan Y., Wang W., Chang K.W., R. Lyu M., Fact-or-Fair: A Checklist for Behavioral Testing of AI Models on Fairness-Related Queries, arXiv:2502.05849 [cs.CL]

(Impett & Offert 2023) Impute L., Offert F., There Is a Digital Art History https://arxiv.org/abs/2308.07464 [cs.CV]

(Jameson 1983), Jameson, F., The Political Unconscious: Narrative as a Socially Symbolic Act, Routledge

(Jay, 1988) Jay, M. Scopic Regimes of Modernity. In H. Foster (Ed.), *Vision and Visuality* (pp. 3-23). Bay Press.

(Kobak & Beres 2019) Kobak, Berens, The art of using t-SNE for single-cell transcriptomics.

(Krizhevsky et Al., 2012) Krizhevsky A., Sutskever I., E. Hinton G., ImageNet Classification with Deep Convolutional Neural Networks (2012) https://courses.cs.duke.edu/compsci527/spring19/papers/Krizhevsky.pdf

(Manovich, 1998) Manovich L., Database as symbolic form, https://manovich.net/index.php/projects/database-as-a-symbolic-form : "Ce que nous avons rencontré ici est un exemple du principe général des nouveaux médias : la projection de l'ontologie d'un ordinateur sur la culture elle-même. Si en physique le monde est fait d'atomes et en génétique il est fait de gènes, la programmation informatique encapsule le monde selon sa propre logique. Le monde est réduit à deux types d'objets logiciels qui se complètent : les structures de données et les algorithmes."

(Manovich, 2009) Manovich L., Cultural Analytics: Visualizing Cultural Patterns in the Era of “More Media”, https://manovich.net/content/04-projects/064-cultural-analytics-visualizing-cultural-patterns/60_article_2009.pdf

(Mikolov *et al*., 2013) Mikolov, T., Chen, K., Corrado, G., & Dean, J. (2013). Efficient Estimation of Word Representations in Vector Space. *arXiv:1301.3781*.

(Mohamed et al. 2020) Mohamed, S., Png, M. T., & Isaac, W. (2020). Decolonial AI: Decolonial Theory as Sociotechnical Foresight in Artificial Intelligence. Philosophy & Technology, 33(4), 659-684.

(Paglen, 2016) Paglen T., Invisibles Image (Your picture are looking at you), https://thenewinquiry.com/invisible-images-your-pictures-are-looking-at-you/

(Radford *et al.* 2021) Radford A., Wook Kim J., Hallacy C., Ramesh A., Goh G., Agarwal S., Sastry G., Askell A., Mishkin P., Clark J., Krueger G., Sutskever I., Learning Transferable Visual Models From Natural Language Supervision, arXiv:2103.00020v1 [cs.CV]

(Rahwan *et al.* 2019) Rahwan I., Cebrian M., Obradovich N., Josh Bongard, Bonnefon J.F., Breazeal C., W. Crandall J., A. Christakis N., D. Couzin I., O. Jackson M., R. Jennings N., Kamar E., M. Kloumann I., Larochelle H., Lazer D., McElreath R., Mislove A., David C. Parkes, ‘Sandy’ Pentland A., E. Roberts M., Shariff A., B. Tenenbaum J., Wellman M., Machine behaviour, https://www.nature.com/articles/s41586-019-1138-y

(Sambasivan et al. (2021) Sambasivan, N., et al. (2021). "Everyone wants to do the model work, not the data work": Data Cascades in High-Stakes AI. CHI 2021

(Simon, 1996) Herbert Simon, The Sciences of the Artificial, (1996), Third Edition.

(Steyerl, 2012) Steyerl, H. (2012). The Wretched of the Screen. e-flux / Sternberg Press

(Van der Maaten & Hinton., 2008) van der Maaten L., Hinton G., Visualizing Data using t-SNE, Journal of Machine Learning Research 9 (2008) 2579-2605. https://www.jmlr.org/papers/volume9/vandermaaten08a/vandermaaten08a.pdf

(Wattenberg *et al*. 2016) — *How to Use t-SNE Effectively, https://distill.pub/2016/misread-tsne/*

(Zhai *et al*. 2023) Zhai X., Mustafa B., Kolesnikov A., Beyer L., Sigmoid Loss for Language Image Pre-Training arXiv:2303.15343v4 [cs.CV]

Illustration 1

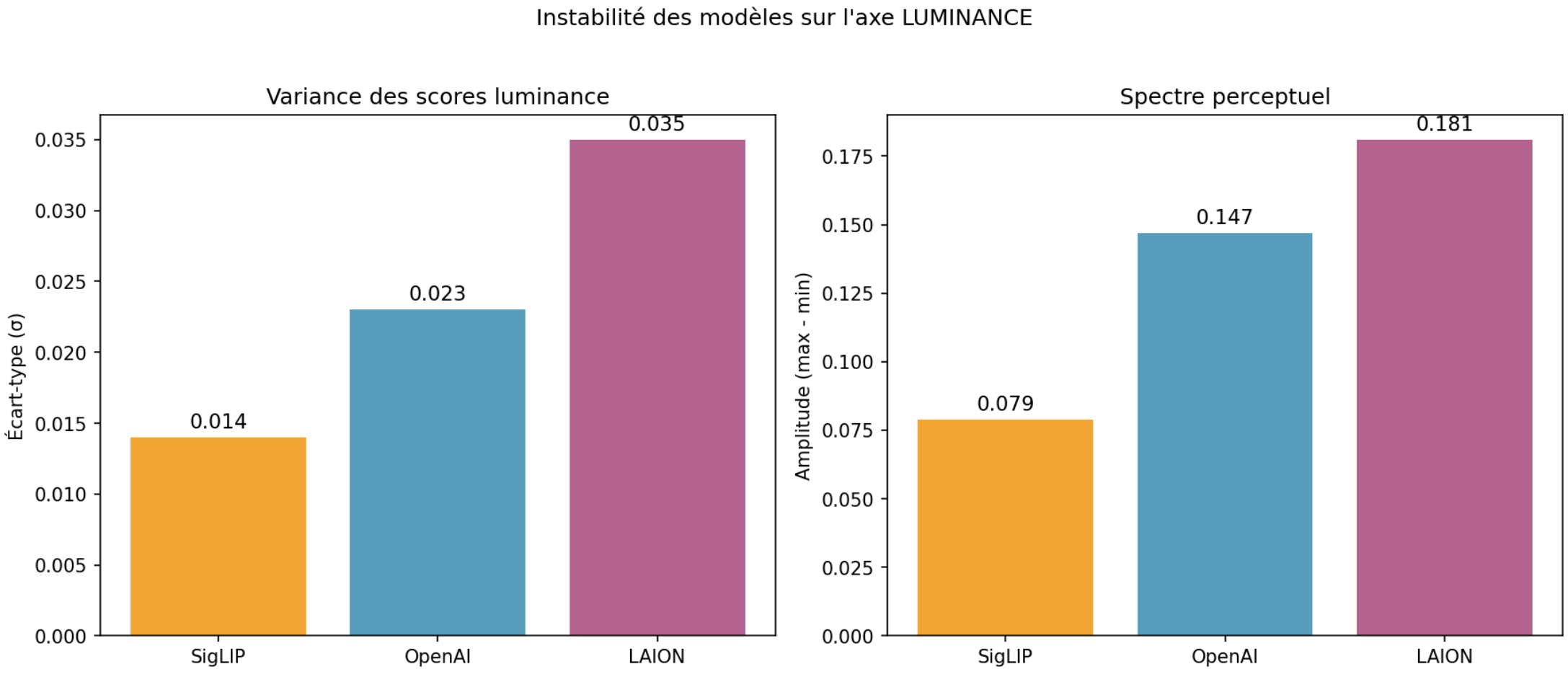

Illustration 2

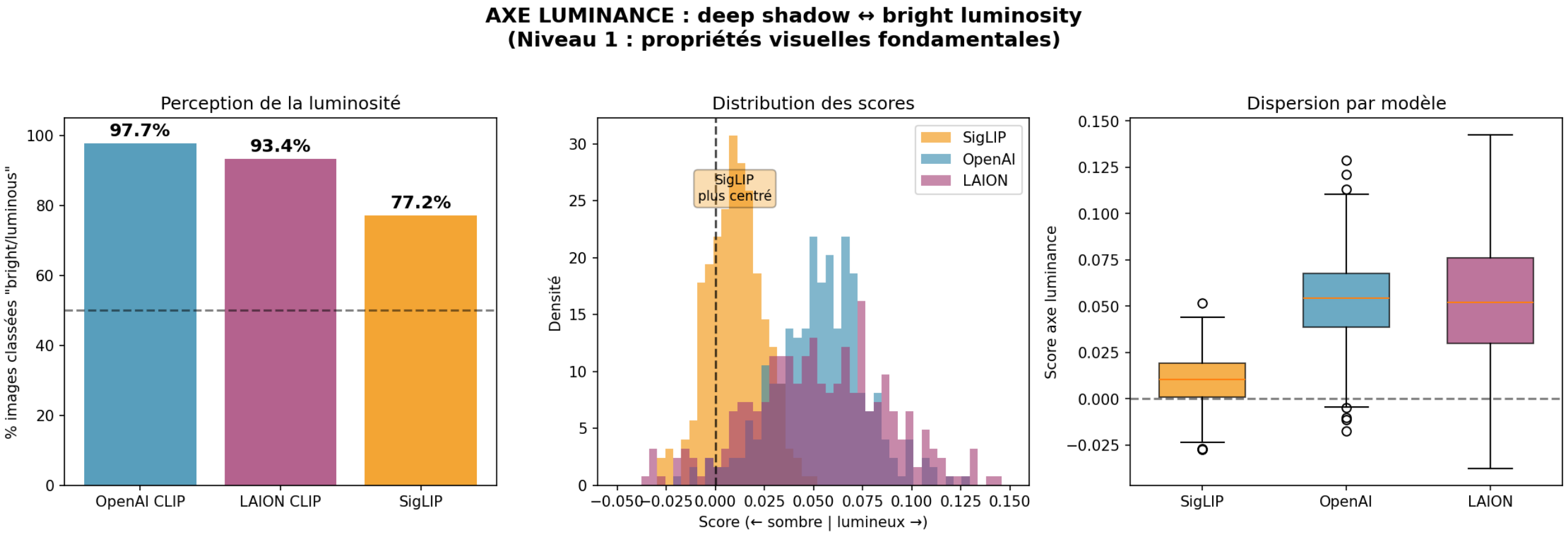

## Illustration 3 : Analyse des corrélations inter-modèles qui met en évidence le cas des néons de Kosuth.

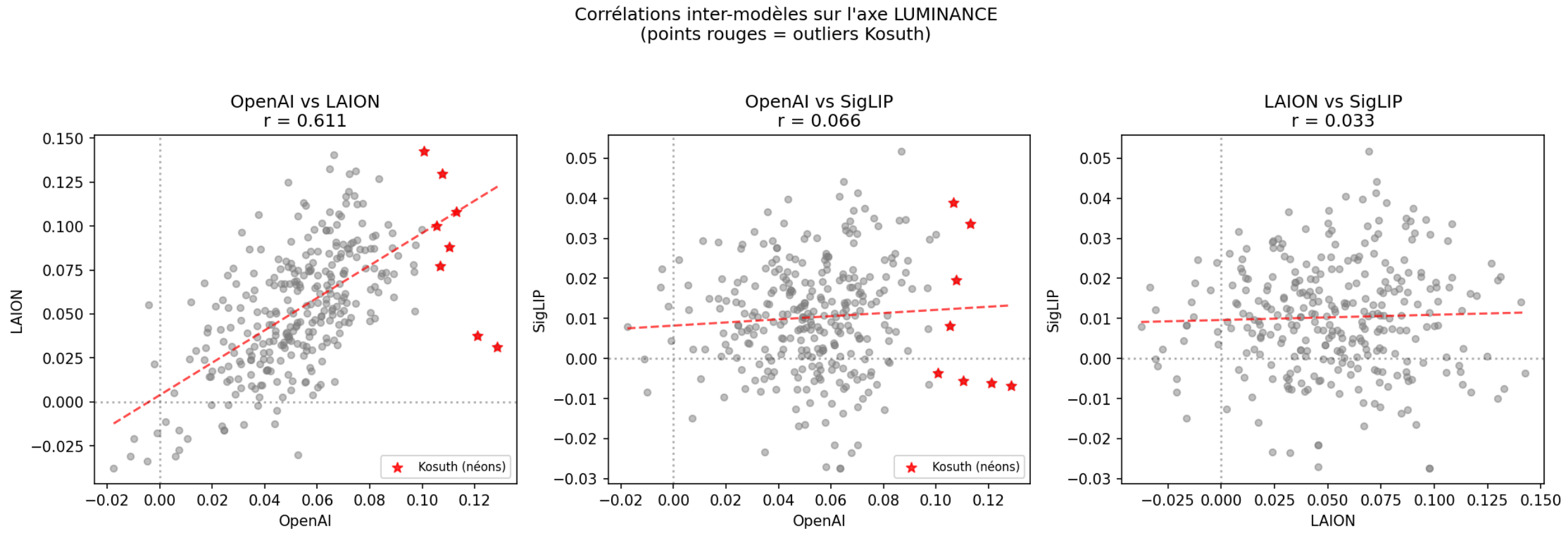

## Illustration 4 : Luminance top 10 des classements selon les modèles

| Rang | OpenAI_Image | OpenAI_Score | LAION_Image | LAION_Score | SigLIP_Image | SigLIP_Score |
|---|---|---|---|---|---|---|
| 1 | kosuth Four Colors Four Words 1965 | 0.1286807656288147 | OADA JAUNE.001 | 0.14256542921066284 | gauguin  Tahiti Arearea or Joyousness Paul... | 0.05173690989613533 |
| 2 | kosuth   Neon (1965) | 0.12107819318771362 | picasso rose 3 | 0.14050264656543732 | picasso les dem davignon | 0.04410869628190994 |
| 3 | gauguin 1 | 0.11306122690439224 | paul signac 2 | 0.13262896239757538 | rembrandt lecon anatomie | 0.041312068700790405 |
| 4 | kosuth Five Words In Green Neon 1965 | 0.11030172556638718 | be93b3f68d48c8a77d6a3451540770cc | 0.1311917006969452 | leger | 0.04054473713040352 |
| 5 | paul signac 4 | 0.10773105174303055 | van gogh Coucher de soleil Montmajour | 0.12971526384353638 | sddefault | 0.03982945531606674 |
| 6 | gauguin   Tahiti I Raro Te Oviri (Under th... | 0.1067085787653923 | paul signac 4 | 0.12968625128269196 | gauguin   Tahiti I Raro Te Oviri (Under th... | 0.03893143683671951 |
| 7 | kosuth Neon Electrical Light English Glass... | 0.10536087304353714 | jean rustin 04 | 0.12675024569034576 | arton192 4250e | 0.03748971223831177 |
| 8 | OADA JAUNE.001 | 0.1006980687379837 | paul signac 3 | 0.1247330754995346 | gauguin Fatata te Miti By the Sea | 0.037001438438892365 |
| 9 | gauguin 2 | 0.09996582567691803 | paul signac 5 | 0.11987284570932388 | léger+ +Women+in+an+Interior+ | 0.036594972014427185 |
| 10 | gauguin 3 | 0.09748867899179459 | van gogh maison jaune | 0.11716759204864502 | kirchner 1 | 0.03499719500541687 |

`

## Illustration 5 :

| Rang | OpenAI_Image | OpenAI_Score | LAION_Image | LAION_Score | SigLIP_Image | SigLIP_Score |
|---|---|---|---|---|---|---|
| 1 | meidner 12 | -0.018 | meidner 12 | -0.038 | rembrandt zelfportret mh840 mauritshuis | -0.027 |
| 2 | meidner 11 | -0.011 | meidner 13 | -0.034 | rembrandt zelfportret mh840 mauritshuis 1 | -0.027 |
| 3 | Hopper night shadows | -0.010 | meidner 3 | -0.031 | bellmer 34 | -0.027 |

## Illustration 6

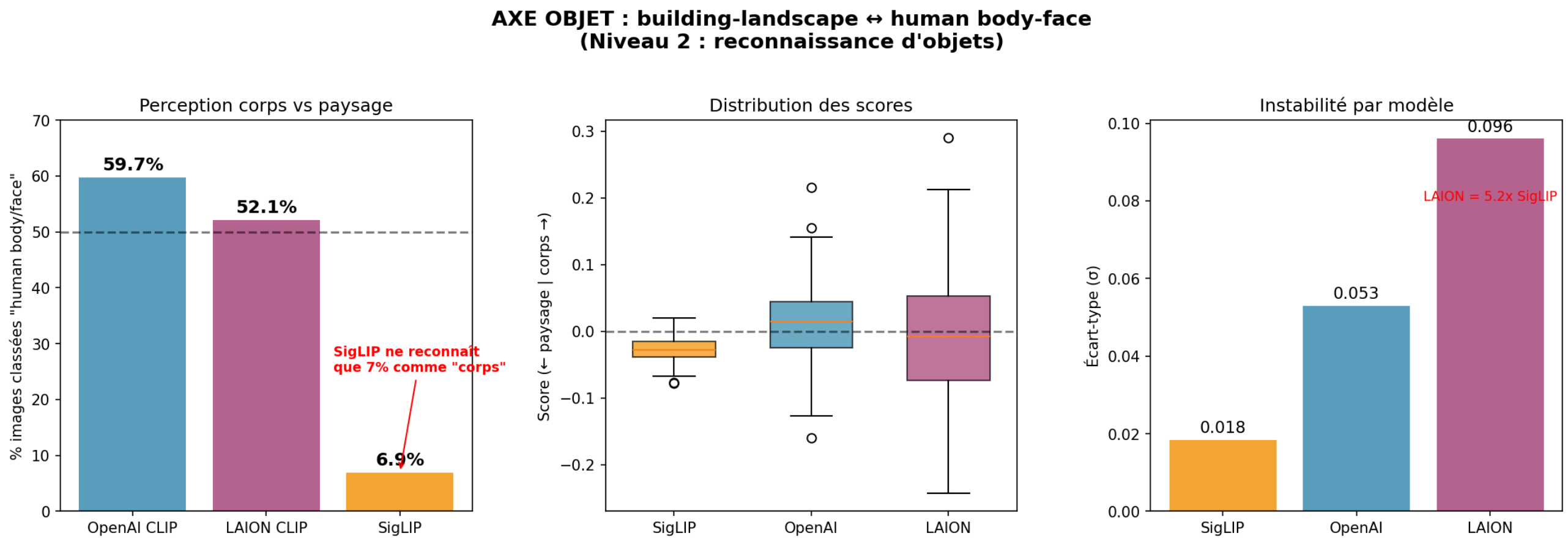

# Illustration 7

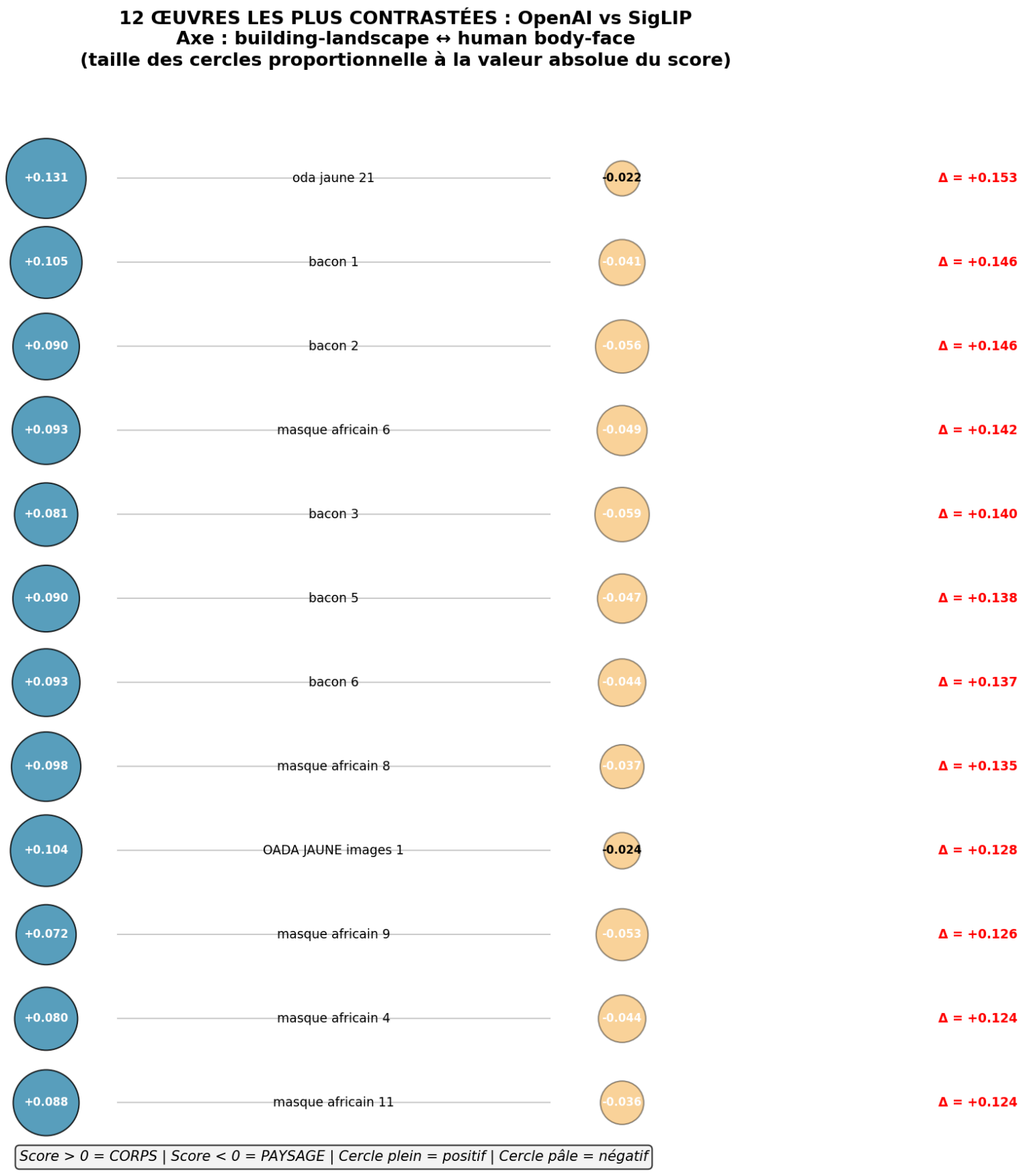

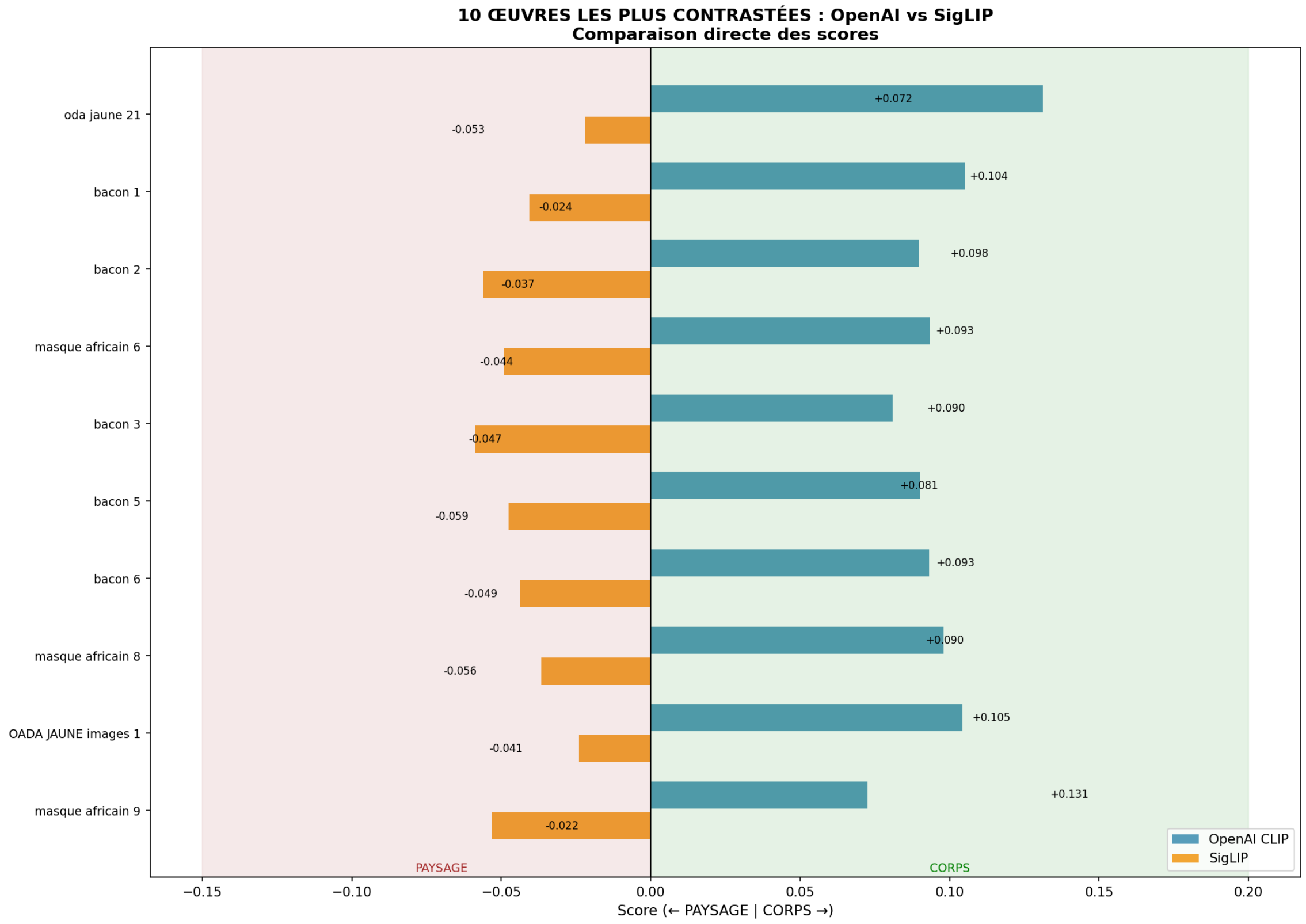

## Illustration 8

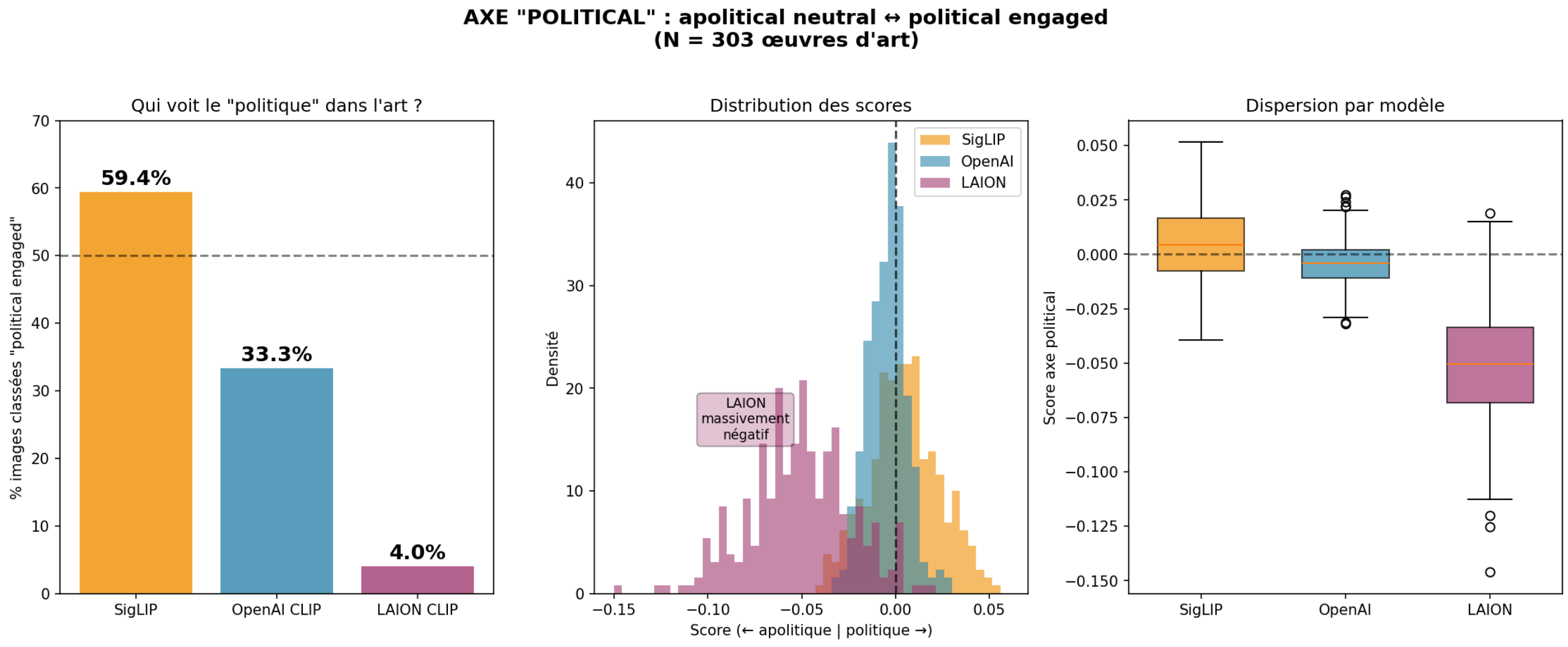

## Illustration 9

| Rang | Œuvres les PLUS engagées | Score | Œuvres les MOINS engagées | Score | | |
|---|---|---|---|---|---|---|
| 1 | Kirchner | Eine Künstlergemeinschaft | +0.0272 | Picasso | Garçon conduisant un cheval | −0.0319 |
| 2 | Kosuth | Neon | +0.0265 | Chéret Autoportrait | −0.0314 | |
| 3 | Larry Clark | +0.0241 | Picasso | période rose (Frère) | −0.0289 | |
| 4 | Otto Dix | Street Chaos | +0.0220 | Masque africain | −0.0279 | |
| 5 | Larry Clark Kids | +0.0220 | Bacon | −0.0264 | | |
| 6 | Kirchner | Milly endormie | +0.0202 | Image abstraite | −0.0255 | |
| 7 | Kirchner | Quatre baigneuses | +0.0180 | Vermeer | L'Astronome −0.0252 | |
| 8 | Larry Clark | Tulsa | +0.0163 Picasso | période rose −0.0251 | | |
| 9 | Russolo | Tower Bridge | +0.0156 | Image non figurative | −0.0231 | |
| 10 | Otto Dix Chaos dans la rue | +0.0137 | Image indéterminé | e −0.0230 | | |

## Illustration 10

| image_index | image_relpath | score_axis | cos_left | cos_right | certainty_mode | certainty |
|---|---|---|---|---|---|---|
| 166 | masque_africain_1.jpeg | -0.0027167052030563354 | 0.13917076587677002 | 0.13645406067371368 | margin | 0.0027167052030563354 |
| 167 | masque_africain_10.jpg | -0.021519705653190613 | 0.1243792474269867 | 0.10285954177379608 | margin | 0.021519705653190613 |
| 168 | masque_africain_11.jpg | -0.01597529649734497 | 0.13670672476291656 | 0.1207314282655716 | margin | 0.01597529649734497 |
| 169 | masque_africain_12.jpg | -0.013767287135124207 | 0.16609834134578705 | 0.15233105421066284 | margin | 0.013767287135124207 |
| 170 | masque_africain_13.jpg | -0.02065497636795044 | 0.1341075301170349 | 0.11345255374908447 | margin | 0.02065497636795044 |
| 171 | masque_africain_14.jpg | -0.0177726149559021 | 0.1373971849679947 | 0.11962457001209259 | margin | 0.0177726149559021 |
| 172 | masque_africain_15.jpg | -0.016980379819869995 | 0.15173572301864624 | 0.13475534319877625 | margin | 0.016980379819869995 |
| 173 | masque_africain_2.jpeg | -0.007254704833030701 | 0.13722266256809235 | 0.12996795773506165 | margin | 0.007254704833030701 |
| 174 | masque_africain_3.jpeg | -0.010861918330192566 | 0.12689296901226044 | 0.11603105068206787 | margin | 0.010861918330192566 |
| 175 | masque_africain_4.jpg | -0.007886126637458801 | 0.13335371017456055 | 0.12546758353710175 | margin | 0.007886126637458801 |
| 176 | masque_africain_5.jpg | -0.027866430580615997 | 0.1515263170003891 | 0.1236598864197731 | margin | 0.027866430580615997 |
| 177 | masque_africain_6.jpg | -0.021440058946609497 | 0.15829887986183167 | 0.13685882091522217 | margin | 0.021440058946609497 |
| 178 | masque_africain_7.jpg | -0.0014204084873199463 | 0.1429780274629593 | 0.14155761897563934 | margin | 0.0014204084873199463 |
| 179 | masque_africain_8.jpg | -0.022211924195289612 | 0.13154031336307526 | 0.10932838916778564 | margin | 0.022211924195289612 |
| 180 | masque_africain_9.jpg | -0.017200753092765808 | 0.13164740800857544 | 0.11444665491580963 | margin | 0.017200753092765808 |

Illustration 11

| Rang | Œuvres les PLUS engagées | Score | Œuvres les MOINS engagées | Score | |
|---|---|---|---|---|---|
| 1 | Schiele Autoportrait chemise rouge | +0.0189 | Michel-Ange | −0.1461 | |
| 2 | Schiele Homme debout | +0.0149 | Image non nommée | −0.1252 | |
| 3 | Larry Clark | +0.0115 | Meidner | −0.1201 | |
| 4 | Russolo Musica | +0.0041 | Picasso | période rose | −0.1125 |
| 5 | Larry Clark Tulsa | +0.0027 | Jean Rustin | −0.1093 | |
| 6 | Oda Jaune | +0.0025 | Jean Rustin | −0.1053 | |
| 7 | Kirchner | +0.0022 | Jean Rustin | −0.1032 | |
| 8 | Gauguin | +0.0021 | Meidner | −0.1020 | |
| 9 | Russolo | +0.0014 | Meidner | −0.1020 | |
| 10 | Gauguin (Tahiti) | +0.0011 | Georges de La Tour | −0.1018 | |

Illustration 12

| Rang | Œuvres les PLUS engagées | Score | Œuvres les MOINS engagées | Score |
|---|---|---|---|---|
| 1 | Masque africain | +0.0517 | Lichtsprayer | −0.0393 |
| 2 | Masque africain | +0.0496 | Jean Rustin | −0.0378 |
| 3 | Picasso cubisme | +0.0488 | Image abstraite | −0.0374 |
| 4 | Masque africain | +0.0461 | Hopper The City | −0.0361 |
| 5 | Picasso masque | +0.0456 | Van Gogh Café de nuit | −0.0357 |
| 6 | Meidner Apocalypse | +0.0446 | Fidus | −0.0347 |
| 7 | Picasso Bouquet | +0.0415 | Oda Jaune  Bellmer | −0.0334 |
| 8 | Russolo Dynamisme | +0.0404 | Picasso période bleue | −0.0321 |
| 9 | Masque africain | +0.0402 | Degas | −0.0311 |
| 10 | Picasso | +0.0400 | Bellmer | −0.0308 |

Illustration 13

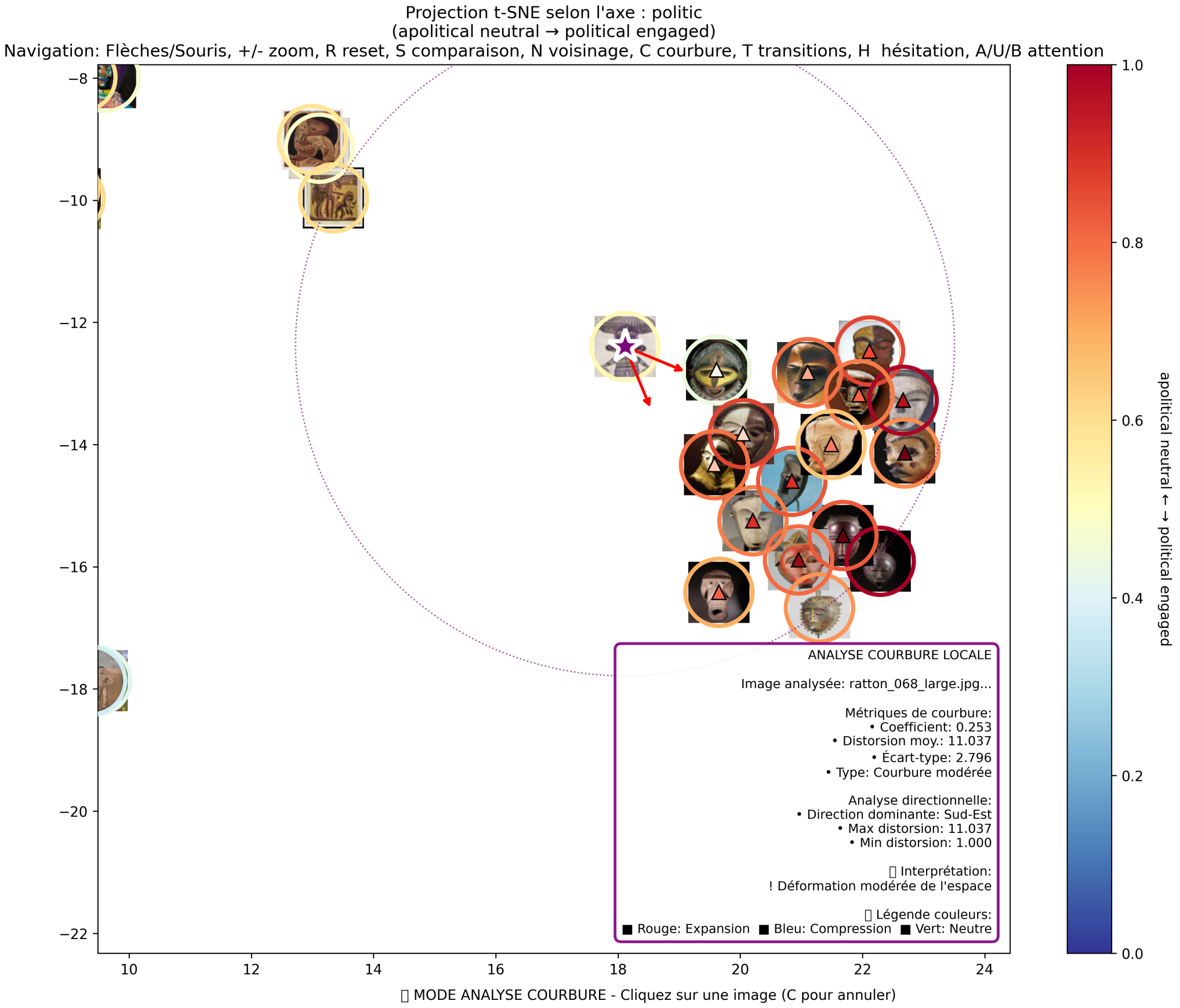
Projection t-SNE selon l'axe : politic
(apolitical neutral → political engaged)
Navigation: Flèches/Souris, +/- zoom, R reset, S comparaison, N voisinage, C courbure, T transitions, H hésitation, A/U/B attention
ANALYSE COURBURE LOCALE
Image analysée: ratton_068_large.jpg...
Métriques de courbure:
• Coefficient: 0.253
• Distorsion moy.: 11.037
• Écart-type: 2.796
• Type: Courbure modérée
Analyse directionnelle:
• Direction dominante: Sud-Est
• Max distorsion: 11.037
• Min distorsion: 1.000
Interprétation:
! Déformation modérée de l'espace
Légende couleurs:
■ Rouge: Expansion ■ Bleu: Compression ■ Vert: Neutre
MODE ANALYSE COURBURE - Cliquez sur une image (C pour annuler)
apolitical neutral ←→ political engaged

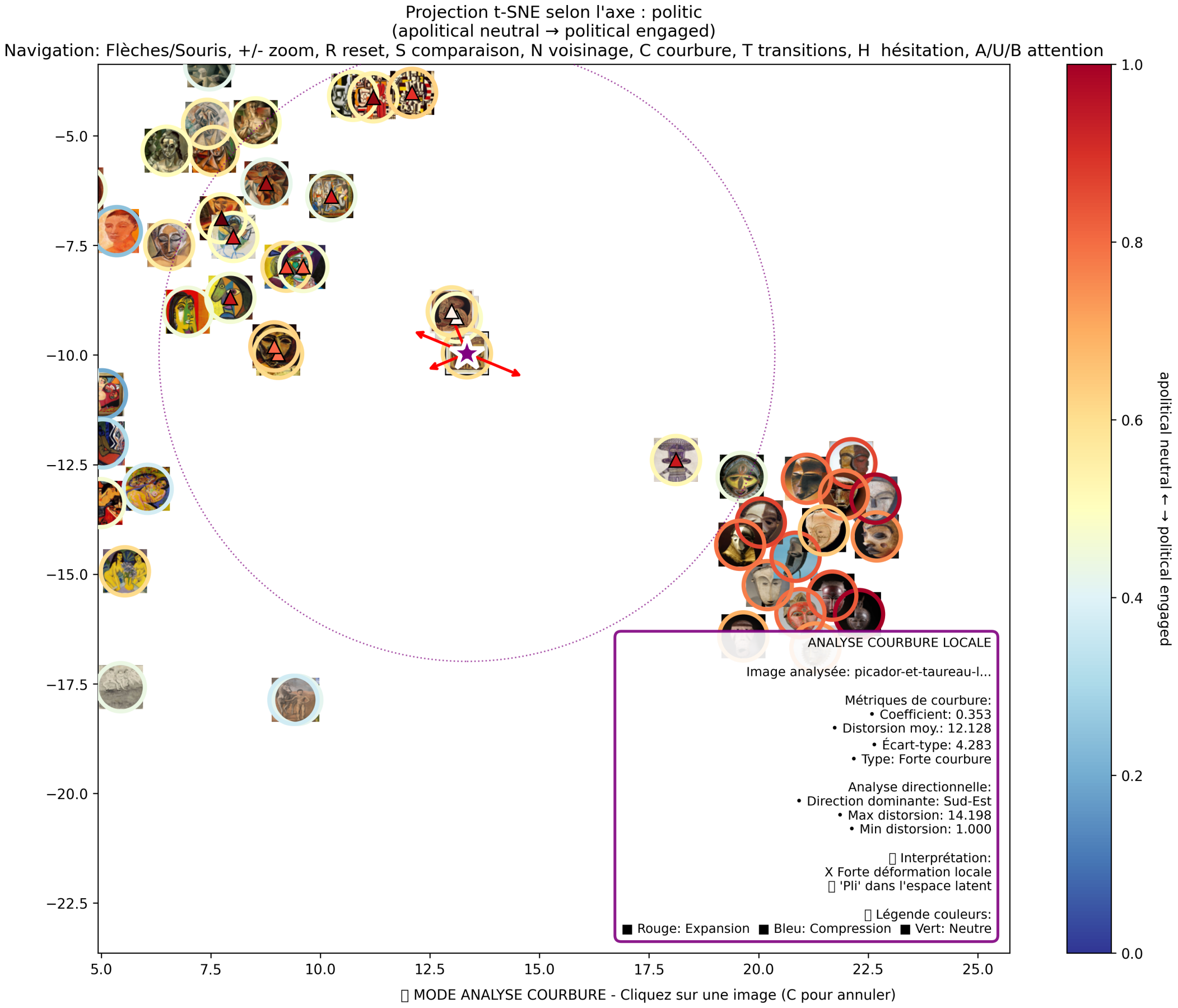
Projection t-SNE selon l'axe : politic
(apolitical neutral → political engaged)
Navigation: Flèches/Souris, +/- zoom, R reset, S comparaison, N voisinage, C courbure, T transitions, H hésitation, A/U/B attention
ANALYSE COURBURE LOCALE
Image analysée: picador-et-taureau-l...
Métriques de courbure:
• Coefficient: 0.353
• Distorsion moy.: 12.128
• Écart-type: 4.283
• Type: Forte courbure
Analyse directionnelle:
• Direction dominante: Sud-Est
• Max distorsion: 14.198
• Min distorsion: 1.000
Interprétation:
X Forte déformation locale
'Pli' dans l'espace latent
Légende couleurs:
■ Rouge: Expansion ■ Bleu: Compression ■ Vert: Neutre
MODE ANALYSE COURBURE - Cliquez sur une image (C pour annuler)
apolitical neutral ←→ political engaged

# Illustration 14

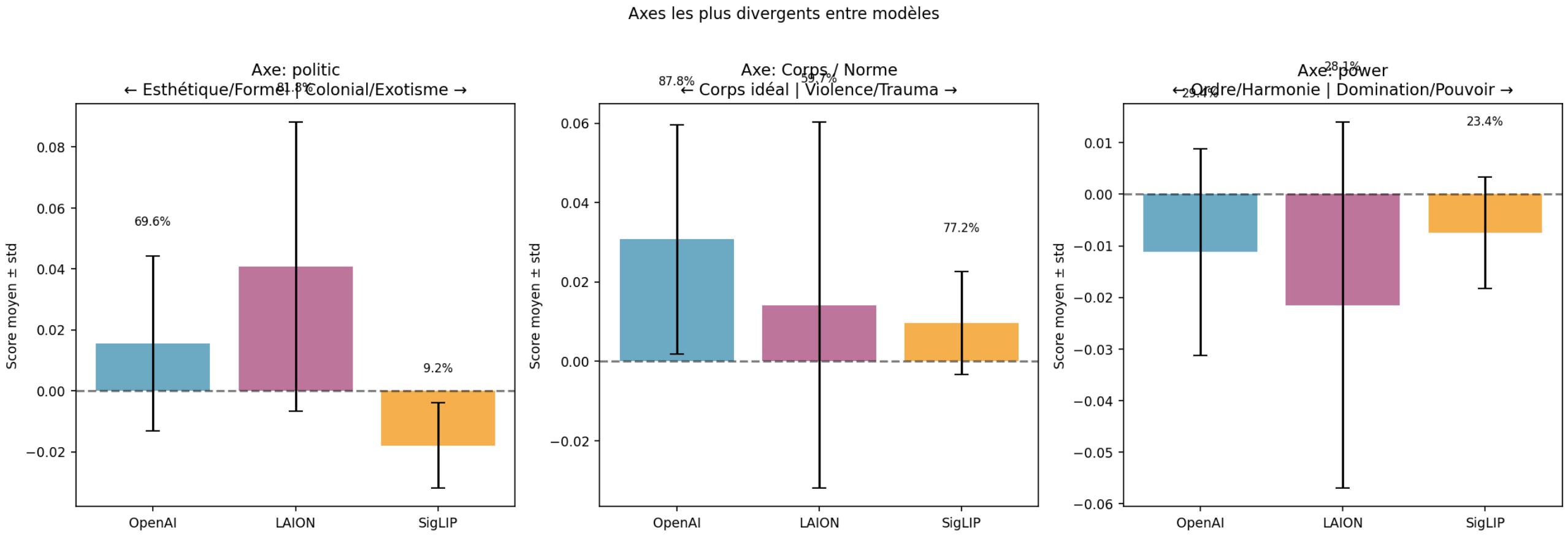